\documentclass{WileyMSP-template}
\UseRawInputEncoding 
\usepackage{graphicx}
\usepackage{color,amssymb,amsthm,bm,bbm,dcolumn}
\usepackage[colorlinks,urlcolor=black,citecolor=black,linkcolor=black]{hyperref}

\newcommand{\ket}[1]{| #1 \rangle }
\newcommand{\bra}[1]{\langle #1 | }

\begin{document}

\title{Quantum simulating the electron transport in quantum cascade laser structures }

\maketitle


\author{Andrea Trombettoni,}
\author{Francesco Scazza,}
\author{Francesco Minardi ${}^{*}$,}
\author{Giacomo Roati,}
\author{Francesco Cappelli,}
\author{Luigi Consolino,}
\author{Augusto Smerzi,}
\author{Paolo De Natale}

\begin{affiliations}
A. Trombettoni \\
\textit{Department of Physics, University of Trieste, Strada Costiera 11, I-34151 Trieste, Italy}
\\
\textit{CNR-IOM DEMOCRITOS Simulation Center, via Bonomea 265, I-34136 Trieste, Italy}

F. Scazza, F. Minardi, G.Roati, F. Cappelli, L. Consolino, A. Smerzi, P. De Natale\\
\textit{Istituto Nazionale di Ottica (CNR-INO), Largo Enrico Fermi 6, 50125 Florence, Italy}

Email: francesco.minardi@ino.cnr.it

F. Scazza, F. Minardi, G.Roati, F. Cappelli, A. Smerzi, P. De Natale\\
\textit{European Laboratory for Non-linear Spectroscopy (LENS), Via Nello Carrara 1, 50019 Sesto Fiorentino, Italy}

F. Minardi \\
\textit{Dipartimento di Fisica e Astronomia, Universit\`a di Bologna, Viale Carlo Berti-Pichat 6/2, 40127 Bologna, Italy}
\end{affiliations}

\begin{abstract} 
We propose to use ultracold fermionic atoms in one-dimensional optical lattices to quantum simulate the electronic transport in quantum cascade laser (QCL) structures. The competition between the coherent tunneling among (and within) the wells and the dissipative decay at the basis of lasing is discussed. In order to validate the proposed simulation scheme, we quantitatively address such competition in a simplified one-dimensional model. 
We show the existence of optimal relationships between the model parameters, maximizing the particle current, the population inversion (or their product), and the stimulated emission rate. This substantiates the concept of emulating the QCL operation mechanisms in cold-atom optical lattice simulators, laying the groundwork for addressing open questions, such as the impact of electron-electron scattering and the origin of transport-induced noise, in the design of new-generation QCLs.

\end{abstract}






\section{Introduction} 
Quantum cascade lasers (QCLs) \cite{Faist:1994,Kohler:2002} are among the most striking examples of how quantum mechanics is nowadays at the basis of everyday-life technology, 
from the implementation of gas-related sensors and radars to cultural heritage investigation \cite{Yao:2012}. 
QCLs are realized by metamaterials, consisting of hundreds of semiconductor layers forming multiple-quantum-well structures, providing an 
extraordinary flexibility and versatility in designing a large number of configurations.
%
%
Since QCLs first demonstration, many fundamental advancements have been achieved regarding their operation capabilities. By improving heat extraction and carrier diffusion, continuous-wave room temperature operation has been achieved for mid-infrared QCLs \cite{beck:2002}, while THz devices have very recently achieved room temperature operation \cite{Bosco:2019,Khalatpour:2020}. 
The emitted power has also been increased, reaching for mid-infrared QCLs the Watt level \cite{bai:2008} and 31\% wall-plug efficiency \cite{wang:2020}. 
Moreover, two fundamental aspects of QCLs emission properties have been clarified. Firstly, the intrinsic linewidth narrowness of QCLs emission has been experimentally demonstrated \cite{Bartalini:2010,Vitiello:2012,Maddaloni:2013bo}. This feature results from the ratio between radiative and nonradiative decay of the laser transition, which for QCLs is unbalanced towards the latter. This relates to a low efficiency but also to a narrow intrinsic linewidth, being the spontaneous emission highly suppressed above threshold. Secondly, the capability of QCLs to generate frequency combs in free-running continuous-wave operation has been demonstrated \cite{Hugi:2012,Burghoff:2014,Faist:2016,Consolino:2021a}. Thanks to the high third-order nonlinearity characterizing the active medium, four-wave mixing couples the modes emitted by Fabry-P\'erot QCLs \cite{Opacac:2019,Burghoff:2020,Khurgin:2020} establishing a fixed phase relation \cite{CappelliConsolino:2019,Singleton:2018}. In order to control and improve further the spectral properties (e.g., the spectral coverage \cite{Riedi:2015} and the frequency/phase noise \cite{Cappelli:2016,Cappelli:2015,Consolino:2019a,Consolino:2021b}), a considerable effort has been undertaken to reduce the group velocity dispersion characterizing the waveguide (Gires-Tournois \cite{Villares:2016} and double-waveguide \cite{Bidaux:2018} approaches). 

However, the quest for the ideal QCL design is a real challenge, made even harder by the complex, 
expensive and time consuming fabrication processes. 
The present day maturity level has been reached by employing advanced computer simulation tools, such as Monte Carlo techniques and non-equilibrium Green's functions \cite{Wacker:2013,Franckie:2018}. They enabled the current performance of QCL devices but, while requiring significant computational resources, they provide an incomplete description of the physical system. Important aspects, such as electron-electron 
interactions or other scattering mechanisms, time-resolved dynamics, and 
coupling with the laser field cannot be fully taken into account, due to their quantum nature, 
despite they are expected to play a fundamental role in the laser performance. 

In this paper we propose an analog quantum simulator consisting of an ultracold Fermi gas trapped in an optical lattice, to study electron transport in QCLs \cite{Morsch:2006,Giorgini:2008,Lewenstein:2012bo}. 
%
The goal is to fully characterize the fundamental QCL structure (i.e.~an array of tunnel-coupled quantum wells) for optimizing QCLs operation. The optical lattice of our quantum simulator reproduces the periodic structure of the semiconductor layers in QCLs, while the laser-assisted tunneling of ultracold fermions emulates electron inter-layer tunneling under an applied voltage. QCL operation is crucially based on dissipation-assisted transport in the injector regions; its performance hinges on a delicate balance between coherent radiative and non-radiative decay processes between narrow energy bands in quantum wells \cite{Faist:2013bo}, whereby efficient electron transport optimizes the laser performance. By quantitatively modelling the tunneling dynamics in the minimal case of 3 levels in each well and non-interacting particles, we study the elementary relations among the key parameters governing QCL lasing, showcasing the potential of atomic quantum simulations for a more thorough understanding of the essential physical processes at play in QCL systems. 
In particular, we report the relative parameter values which maximize the current and the population inversion within a 3-level model structure.

\section{Quantum simulation of relevant physical processes in QCLs}
\label{PhysProc}


An analog quantum simulator is a physical platform whose dynamics reproduces the evolution of a theoretical quantum model that cannot be solved by means of classical numerical approaches. The reason is that complex quantum models evolve in exponentially large Hilbert spaces that would require equally large computational 
matrices that cannot be managed by the most powerful classical computers available today or in the foreseen future. The quantum model is supposed to capture the essential physics of a real quantum system realized in the laboratory.
The main desired characteristics of analog quantum simulators are their scalability and their resilience against noise and decoherence.
A further crucial ingredient is the possibility to accurately tune the
parameters of the experimental platform, -- such as carrier-carrier interactions, tunnel couplings, disorder strength and temperature -- 
to address the largest class of relevant dynamical regimes. 
%
Quantum simulating electron transport in real quantum-engineered devices 
such as QCLs may provide new insights into the quantum mechanical effects 
governing dissipative electron transport within QCL active regions. 
So far, numerical simulation of QCLs in both the 
mid-infrared and THz regions provided several insights into 
transport and gain mechanisms \cite{Jirauschek:2014,Franckie:2018,Franckie:2020}. 
However, further optimization of QCL design requires to include in the simulations many other relevant mechanisms that can affect laser operation, such as
electron-electron interactions \cite{Yang:1985} and the unavoidable presence of disorder \cite{Krivas:2015}. Furthermore, for 
high-resolution spectroscopy applications, it is necessary to minimise both frequency and amplitude noise of the emitted radiation \cite{Borri:2019a,Galli:2013b,Maddaloni:2013bo,Villares:2015c,Galli:2016a,Consolino:2018a,Consolino:2020,Gabbrielli:2021a}. 
%
An atomic quantum simulation of electron transport in QCL structures can give direct insight in the modeling of: 
\begin{itemize} 
\item{\it Electron-electron scattering --} This has been argued to be one of the most important scattering mechanisms \cite{Kohler:2001}, so far 
taken into account via mean-field approaches (Hartree approximation) or 
simplified treatments of the two-body interaction self-energy 
\cite{Bonno:2005,Jirauschek:2010,Winge:2016}. Of special interest is the dependence of the interaction effects on temperature.
In cold atom quantum simulators, the interaction can be controlled by means of magnetic Feshbach resonances, 
which allow the study of both weakly and strongly interacting regimes 
\cite{Chin:2010}, while the temperature can be independently adjusted and accurately measured. 

\item{\it Disorder --} In QCLs an important symmetry-breaking effect between periods is induced by fabrication tolerances, which can also be introduced on purpose to obtain broadband lasing \cite{Gmachl:2002}.  
In ultracold atomic systems, the impact of disorder on transport can be studied by superimposing speckle patterns or incommensurate optical lattices 
on the primary lattice potential \cite{Roati:2008}. In this way, the disorder strength and 
correlation length are fully controllable. 

\item{\it Real-time dynamics --} Typically, QCL models solve 
for the periodic steady state of the system, calculating scattering 
matrix elements for individual values of the applied bias voltage 
\cite{Jirauschek:2014}. In contrast, time-resolved simulations may require 
calculation of the matrix elements at every time step, resulting in much 
more time-consuming computations. Cold atoms could provide 
direct information, since it is possible to visualize their dynamics and the 
relative state populations in real-time by direct imaging of the relevant atomic states. 
In this context, a numerically efficient simplified density matrix approach for simulations over extended time and propagation length scales 
has been developed \cite{Tzenov:2016}. On the other hand, a time-resolved 
full-coherence density matrix model is available \cite{Gordon:2009, Kirsanskas2018, Markmann2021}. They can both serve as validation tools for cold atoms simulations. 

\item{\it Transport-induced QCL noise --} A relevant contribution to the laser noise is given by the electron transport itself. This noise appears to be intrinsic to the devices operation 
\cite{Bartalini:2010,Tombez:2013a} and can be at least partially explained with the filling and emptying of impurity states in injector regions \cite{Yamanishi:2014}. Additional noise might come from electric field domains forming when QCLs operate in a regime of negative differential conductivity \cite{Dhar:2014}. This kind of noise, manifesting in the lasers emission frequency and amplitude, is mirrored in the bias current, whose fluctuations could be theoretically investigated and experimentally addressed with tunable parameters and an expected good statistics in a cold atom simulator. 
\end{itemize}



\section{The simulator scheme}
\label{SIMUL}

The proposed simulation of electron transport in QCLs is based on ultracold fermionic atoms tunneling through a one-dimensional optical lattice. Periodic lattice potentials with adjustable well spacing are routinely generated by interfering two laser beams under a suitable angle.  
The periodic potential generated by the resulting standing-wave pattern consists of an array of equidistant wells, whose height is simply tuned by setting the laser intensity. In this way, ultracold atoms can be trapped in individual quasi-two-dimensional layers, separated by potential barriers. This geometry reproduces that of QCL heterostructures, in which two-dimensional semi-conductor layers are tunnel-coupled through thin insulating barriers. Interactions between internal atomic states, parametrized by the $s$-wave scattering length, can be controlled via Feshbach resonances. In this way it is possible to smoothly tune the gas from weakly to strongly interacting by the application of an external homogeneous magnetic field. 

\begin{figure}[!htbp]
\begin{center}
    \includegraphics[width=\textwidth]{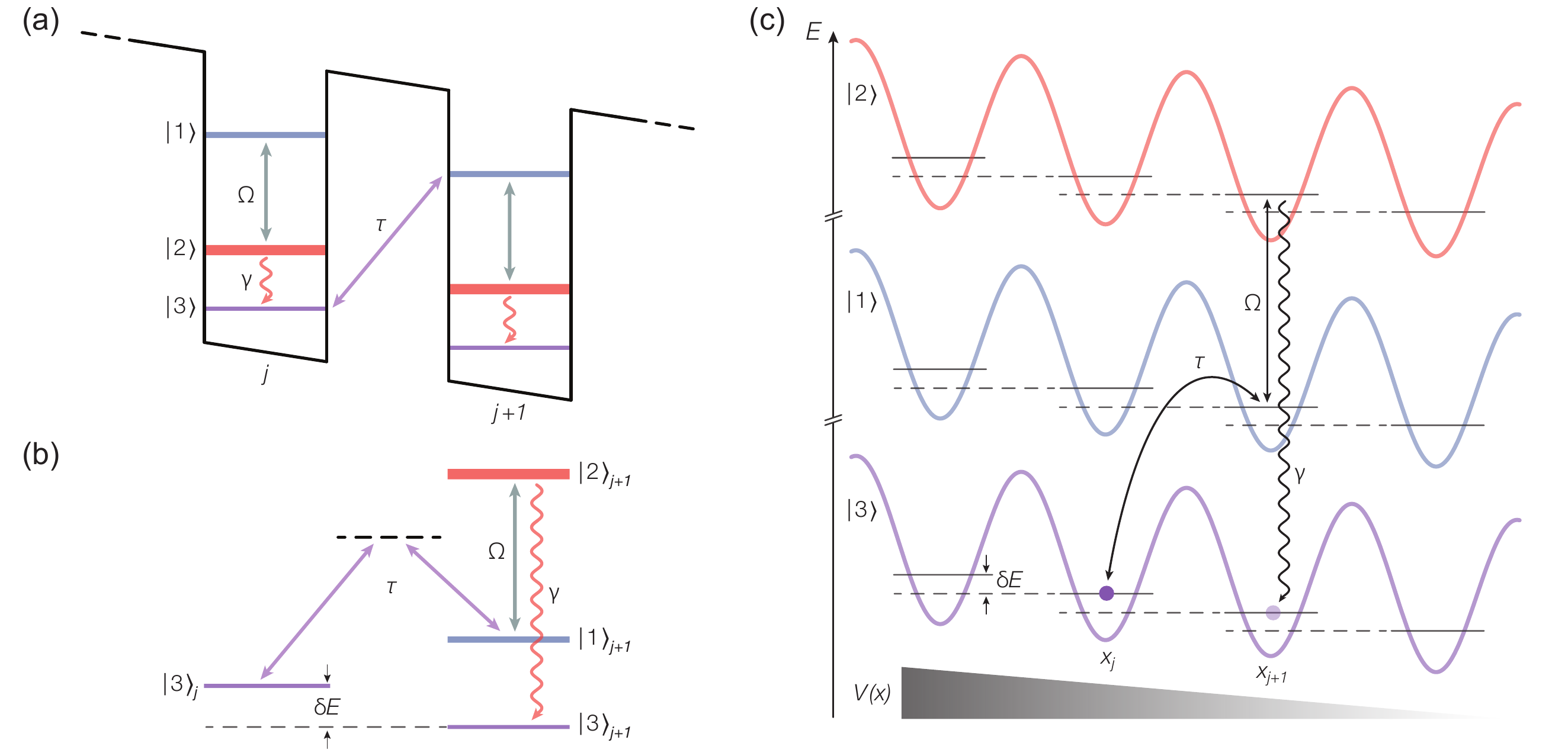}
    \caption{Cold-atom implementation of dissipation-assisted directional transport, driven by a combination of coherent (straight arrows) and incoherent (wavy arrows) couplings in a tilted periodic potential.
    (a) Illustration of the QCL-like laser-assisted transport mechanism media within a minimal 3-level model. A first laser field couples coherently with strength $\tau$ two states localized in neighboring wells $j$ and $j+1$, labeled as $\ket{1}$ and and $\ket{3}$, driving the tunneling in a tilted array of quantum wells, which would otherwise be suppressed. In turn, state $\ket{1}$ is coherently coupled with strength $\Omega$ to the lower laser level $\ket{2}$ localized in the same well, which spontaneously decays onto state $\ket{3}$ with rate $\gamma$.
    %
    %
    (b) Detailed implementation of the laser-assisted tunneling process with cold atoms in a one-dimensional lattice. Here, $\ket{1}$ (blue), $\ket{2}$ (red) and $\ket{3}$ (violet) refer to different internal states of the atom. In the sketch, atoms on site $j$ are transferred to site $j+1$ by a two-photon Raman-assisted process, changing their internal state from $\ket{3}$ to $\ket{1}$. The state $\ket{1}$ is optically pumped back to state $\ket{3}$ through a laser resonant with the $\ket{1} \rightarrow \ket{2}$ transition. 
    (c) In a full one-dimensional lattice realization, unidirectional atomic transport is driven by the combination of the three couplings $\tau$, $\Omega$ and $\gamma$.
    }
    \label{fig:transport_cartoon}
\end{center}
\end{figure}

\begin{figure}[!htbp]
\begin{center}
    \includegraphics[width=1.0\columnwidth]{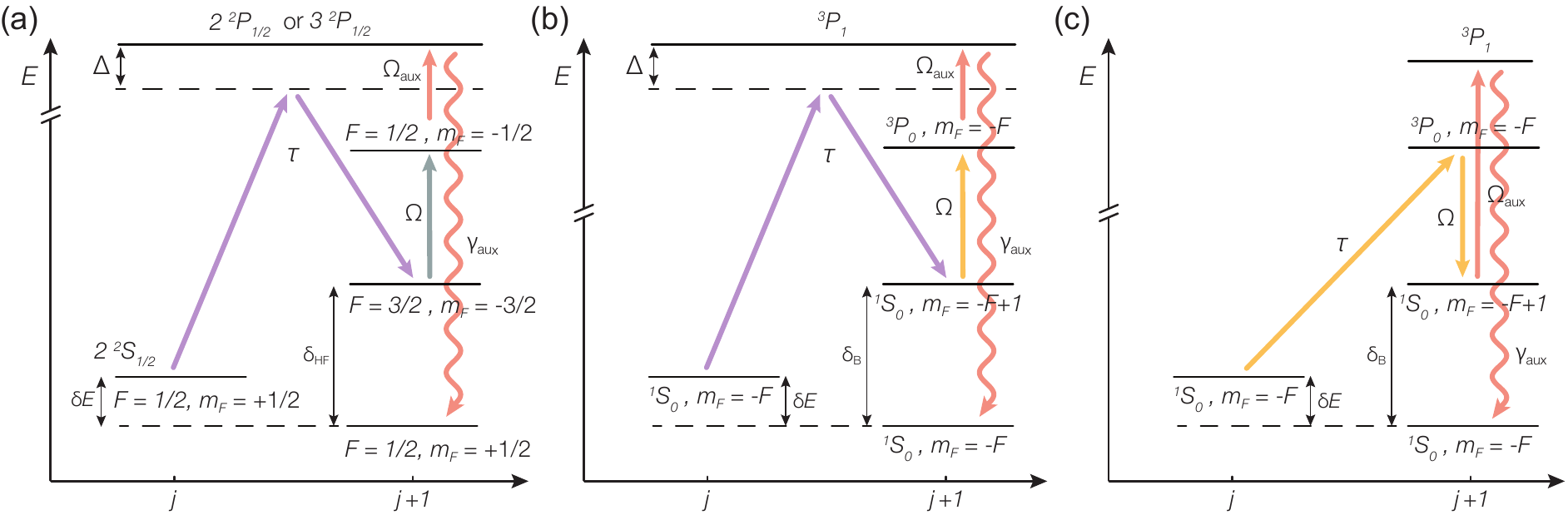}
    \caption{Detailed proposed implementation of the required tunable couplings for: (a) $^6$Li atoms, or (b)-(c) fermionic two-electron atoms, such as $^{87}$Sr or $^{173}$Yb. While in (a) and (b) a two-photon Raman coupling is used to engineer internal state-changing tunneling processes with amplitude $\tau$, in (c) a single-photon optical \emph{clock} laser coupling is exploited. 
    In order to tune the dissipative, spontaneous emission rate $\gamma$, the naturally long-lived lower laser level $\ket{2}$ is optically coupled to a short-lived higher-lying state with a tunable strength $\Omega_\mathrm{aux}$, quenching its lifetime 
    $\gamma \simeq \gamma_\mathrm{aux}\, \Omega_\mathrm{aux}^2/(\gamma_\mathrm{aux}^2+4\Delta_\mathrm{aux}^2)$, where the auxiliary laser is detuned from resonance by $\Delta_\mathrm{aux}$ \cite{Leibfried2003,Nemitz2016}.
    }
    \label{fig:atomic-scheme}
\end{center}
\end{figure}

Here, we consider using different internal states of the atoms to mimic the different levels in the quantum wells of QCL heterostructures. 
In particular, we propose to simulate a 3-level QCL structure, where each of the inter-level coupling strengths can be tuned (see Fig.~\ref{fig:transport_cartoon}). In particular, the relevant couplings between levels are provided by different laser-assisted processes: (i) inter-well tunneling (i.e., injection and extraction) 
is engineered through laser-assisted tunneling in a tilted lattice potential; (ii) coherent coupling (analogous to stimulated emission) is provided by 
coherently driving the transition between the two analog laser levels; (iii) the dissipative coupling (analogous to non-radiative depletion) 
is emulated by tuning the spontaneous decay rate of the lower laser level, via optical quenching to a short-lived 
higher-energy atomic state. Possible atomic implementations of these couplings are sketched in Fig.~\ref{fig:atomic-scheme}.
Alternatively, dissipation could be introduced by adding an interacting atomic bath that surrounds the lattice-trapped atoms and is unaffected by 
the lattice potential. For a bosonic medium, this would emulate the presence of phonons and phonon-assisted nonradiative decay of the QCL laser levels \cite{Faist:2013bo}.

A tunable linear magnetic-field or optical gradient can produce the analogue of the bias voltage imposed at QCLs electric contacts. This external potential introduces a uniform offset between the energy minima of adjacent wells, and it is  key to obtain uni-directional transport in the system. For weak gradients and in the absence of driving, interactions or disorder, the atoms perform Bloch oscillations, and no net current along the lattice is observable. A strong inter-well offset larger than the width of the lowest lattice band localizes the atomic wavefunctions into few lattice sites, inhibiting Bloch oscillations. The tunneling of fermions between neighboring lattice sites along this tilted washboard potential is then restored by a direct laser coupling between the localized Wannier-Stark states centered on adjacent sites 
(similarly to what implemented in Ref.~\cite{Beaufils:2011}). In particular, directional transport can be driven by dissipation through spontaneous photon emission, with a scheme reminiscent of two-photon Raman sideband cooling \cite{Parsons:2015} (see Fig.~\ref{fig:transport_cartoon}). Each transport step ends with the spontaneous emission of a photon after the atom has moved by one lattice spacing. No subsequent transport step can occur in the absence of such dissipative process.  
Furthermore, in principle, some noise can be added on the tilting gradient emulating the noise on the QCL current driver. 


%
%

The described setup is ideal to investigate the (combined) 
effects of disorder and interactions, which are thought to play a key role in QCLs performance. In the proposed experiment, both interactions and disorder can be introduced and controlled in a continuous fashion. 
Interactions between fermions can 
be tuned through Feshbach resonances, from zero to strong, accessing the regime where many-body correlations govern the transport dynamics
\cite{Chien2015}. Disordered potentials can be generated by additional high-resolution optical potentials. 
By playing with the geometrical arrangement of the generating laser beams, both ``in-plane'' and longitudinal randomness can be introduced. 
The ``in-plane'' disorder can be realized by shining on the atomic 
planes speckle or binary (Poissonian) disorder patterns \cite{Morong2015,Choi2016}. 
They can be tailored through digital micro-mirrors devices (DMD), 
exploring both static and dynamical disorder. These potentials can 
be imprinted on a spatial scale comparable with the 
mean inter-particle distance to reach the regime where quantum effects 
are dominant. 

In QCLs, longitudinal disorder ({\it i.e.} along the 
direction of propagation) arises from the 
imperfect growth of the layers during the molecular-beam epitaxy deposition. 
The performance of QCLs is strongly affected both by 
the relative arrangement of the layers and by the background doping level. 
To simulate such effect with ultracold fermions, a quasi-periodic lattice can be created by overlapping 
a weak lattice, incommensurate to the main one, 
that shuffles the energy minima, which become
non-periodically modulated (periodicity is restored only at the very long length scale of the beating between 
the two lattices, i.e. several microns). This quasi-periodic system 
displays a transition from extended to localized states analogous to 
the Anderson transition, already in one dimension for a non-interacting gas 
\cite{Roati:2008}. The combination of disorder and interactions modifies 
the dynamics along the lattice, mimicking the transport of electrons in 
the “non-perfect” QCL heterostructures. 

To reproduce even more closely the real QCL structure, more exotic light potentials can be designed. As an example, mesoscopic lattices can be implemented, as recently demonstrated in \cite{Lebrat:2018}. 
They may include a series of few thin optical barriers, imprinted onto the atoms with high spatial resolution by a DMD. By controlling and tuning their number, shape and relative 
separation, it is possible to study the dynamics in 
different conditions to eventually optimize transport along the lattice. The injection of an external current could be simulated by inserting the mesoscopic lattice in between two atomic reservoirs \cite{Lebrat:2018}, and rigidly moving the potential at constant velocity \cite{Kwon:2020}.

\section{Simulator model and experimental implementation}
The ultracold gas simulator, engineered as discussed above, allows to investigate a master equation of the form
\begin{equation}
\dot{\rho}=-\frac{i}{\hbar} [H,\rho] + {\cal L}[\rho],
\label{eq:rho}
\end{equation}
where $H$ is the effective Hamiltonian governing the coherent 
motion of fermions in the lattice and ${\cal L}[\rho]$ is the Lindbladian describing the spontaneous emission. To fix the notation, in this Section
we denote by the index $i$ the wells, {\it i.e.} the minima 
of the periodic potential, and by 
$\ket{j,1}$, $\ket{j,2}$, $\ket{j,3}$ the kets denoted 
respectively by $\ket{1}$, $\ket{2}$, $\ket{3}$
in the well $j$ in Fig.~\ref{fig:transport_cartoon}(a). 

The Hamiltonian entering Eq.~(\ref{eq:rho}) refers, in general, to a spinful multi-component Fermi gas and the corresponding master equation (\ref{eq:rho}), once that the appropriate Lindblad operator ${\cal L}$ is specified, is of course very hard to solve, due to the 
fact that the Hilbert space grows exponentially with the number of sites. 
Instead, the time-dynamical solution of Eq.~(\ref{eq:rho}) 
would be provided by the ultracold quantum simulator which we propose to experimentally realize.

\subsection{Simplified Model Hamiltonian}

The central problem faced by all quantum simulators is the certification of their results. 
There are not available in the literature quantum error-correction protocols allowing to preserve the fidelity of the simulator quantum state with the state of the system that must be simulated. The only possible
strategy is to benchmark the experimental results in dynamical regimes that are accessible to a theoretical
numerical analysis. To this purpose, it is therefore important to 
develop solvable toy-models that can capture some essential features of the full quantum system that must be simulated. 
It is important to emphasize that these toy-models, however, will not merely act as simple test-beds,
but can provide useful insights on the competition 
of various time scales and energies of the real complex system 
in terms of a few dimensionless control parameters. For these reasons, we will develop and study in this Section 
a one-dimensional, non-interacting model of the quantum electron transport in QCL structures. 
Our model consists of non-interacting particles that can tunnel between weakly-coupled adjacent wells, 
each containing three non-degenerate energy levels. We study the interplay between the intra- and inter-wells coherent dynamics
in presence of dissipation, which leads to an effective dynamical phase diagram containing quantum-Zeno frozen islands 
surrounded by coherent transport regions. Our model is similar to a previous model suggested in the literature 
where 
a one-body density matrix was written in the two-level, or in the so-called {\it full} approximation (see Chapter 12 in Ref.~\cite{Faist:2013bo}),
and could not therefore capture the competition among tunneling and internal dynamics that we study below.
The simplified model can be experimentally simulated by 
using an atomic Fermi gas (e.g., ${}^6$Li atoms) in the setup illustrated in Figs.~\ref{fig:transport_cartoon} and  \ref{fig:atomic-scheme}. Our toy model does not include effects such as interparticle 
interactions, defects, and interactions with transverse degrees of freedom, 
whose full characterization is challenging for numerical studies,
but illustrates the competition between tunneling, coherent oscillations and spontaneous emission in a simple, and experimentally accessible scenario.

The Hamiltonian $H$ has the general form
\begin{equation}
H=H_{\tau}+H_{\Omega}.
\label{eq:Ham}
\end{equation}
$H_\tau$ describes the tunneling from a well to the next one and, more precisely, 
the tunneling from the level $\ket{j,3}$ in the $j$-th 
well to the level $\ket{j+1,1}$ in the $(j+1)$-th well 
\footnote{Notice that for tunneling between wells of an optical lattice the notation with the minus sign in eq.~(\ref{eq:H_tunn}) is often used. Such a different sign can be reabsorbed, and for the initial conditions considered in the following it can be easily seen that the different sign of the parameter $\tau$ with respect 
to the parameter $\Omega$ in eq.~(\ref{eq:H_Omega}) does not modify the results for the population presented in Fig.~\ref{fig1_and}.}: 
\begin{equation}
H_{\tau}=\tau\sum_{j} (c^\dag_{j,3} c_{j+1,1}+h.c.),
\label{eq:H_tunn}
\end{equation}
where 
$c^\dag_{j,\alpha}$ is the fermionic operator 
creating a particle in the level $\ket{j,\alpha}$, i.e. in the state $\alpha$ (with $\alpha=1,2,3$) of well $j$. Instead, $H_\Omega$ describes the coherent coupling from $\ket{j,1}$ to $\ket{j,2}$ in each well $j$:
\begin{equation}
H_{\Omega}=\Omega \sum_{j} (c^\dag_{j,1} c_{j,2}+h.c.).
\label{eq:H_Omega}
\end{equation}

To simulate disorder, on-site energy terms may be added, in order to have wells with different zero point energies. Moreover, it should be noticed that the Hamiltonian (\ref{eq:Ham}), with (\ref{eq:H_tunn}) and (\ref{eq:H_Omega}) can be seen, in the language of electrons hopping in a lattice, as a 
"spinless" Hamiltonian, in the sense that  a single fermionic species is present and at most one fermion can occupy the state $\ket{j,\alpha}$. 
In principle, a further degree of freedom can be added, to obtain 
a "spinful" effective Hamiltonian and to accomodate even contact interactions of two fermions in the same $\ket{j,\alpha}$ state. 
With atomic gases both local and non-local interactions \cite{Lahaye:2009} can be engineered; however, in the simplified  model presented here, we do not consider the effect of interactions, paralleling the simplified density matrix models used in the QCL literature \cite{Faist:2013bo}. 
It is 
important to note that the simplified Hamiltonian (\ref{eq:Ham}) assumes that all fermions do occupy a single motional band in the lattice. However, for typical values of $\tau$ (see Section \ref{comments}), atoms confined in the lattice are not well described by the Lamb-Dicke regime, and spontaneous photon emission leads to a finite population of excited motional bands. These remain uncoupled in absence of interactions, but would need to be explicitly taken into account for describing the interacting problem.

Our main goal is to explore  
competition between the coherent processes ($\propto \tau, \Omega$) and the 
spontaneous dissipative terms. Needless to say, the understanding of the "spinless" case provides the basis for more complete models needed to quantum simulate mechanisms relevant in QCL structures.  
It is now necessary to specify how the Lindbladian operator can be expressed. To describe spontaneous emission, see 
Figs.~\ref{fig:transport_cartoon} and  \ref{fig:atomic-scheme}, following the toy model discussed before, we choose 
\begin{equation}
{\cal L}[\rho]=\sum_j \left( \gamma_j L_j \rho L_j^\dag - (\gamma_j/2) \{L_j^\dag L_j, \rho \}\right),
\label{LIND}
\end{equation} 
with 
\begin{equation}
L^{(j)}=c^{\dag}_{j,3}  c_{j,2}. 
\end{equation}
Since we do not consider the effect of inhomogeneities
between different wells, we will consider all the $\gamma_i$'s equal by putting $\gamma_i \equiv \gamma$. In the next Section we comment on the experimental values of the parameters, and their tunability within the schemes illustrated in Fig.~\ref{fig:atomic-scheme}.

\subsection{Parameter values in experimental implementations}
\label{comments}
The simplified model introduced in Eqs.~(\ref{eq:rho}), (\ref{eq:Ham}) and (\ref{LIND}) features 
three main parameters: 
the effective strength of the 
tunneling term 
between adjacent wells, $\tau$; the
coherent coupling  
between a pair of levels in each well, $\Omega$; and the
the incoherent leakage of particles into the lowest level of each well, $\gamma$. 

In Section \ref{SIMUL} we have shown schemes of a simulator that can be used for alkali fermionic atoms, such as lithium, or for alkaline-earth-like fermionic atoms such as ytterbium or strontium. Since in the next Section we are going to present estimates of physical quantities with varying $\tau$, $\Omega$ and $\gamma$, it is useful to briefly discuss how these parameters can be effectively changed in the different schemes.

Firstly, let us refer to the scheme in panel $(a)$ of Fig.~\ref{fig:atomic-scheme}, valid for fermionic alkalis such as $^6$Li. The parameter $\Omega/h$ (where as usual $h=2\pi\hbar$) can be varied in the range $\sim 100$\,Hz - $10$\,kHz, while the parameter $\tau/h$ 
likely can be adjusted between $\sim 10$\,Hz - $100$\,Hz, considering reasonable laser intensities and lattice depths, while avoiding excessive spontaneous emission and associated heating from the intermediate state in the two-photon scheme. 
On the other hand, the spontaneous emission rate $\gamma$, induced by optically coupling the naturally long-lived level $\ket{2}$ to short-lived higher-lying states, can be varied from $0$ to tens of kHz \cite{Leibfried2003,Nemitz2016}.

In the scheme from panel $(b)$ of Fig.~\ref{fig:atomic-scheme}, referring to e.g. Yb or Sr atoms, thanks to the presence of narrow transitions reducing the undesired spontaneous emission, the laser-assisted tunneling amplitude may reach values $\tau/h \sim 100$\,Hz \cite{Gerbier:2010}, while  the tunability range for $\Omega$ and $\gamma$ is similar to alkali atoms in panel  $(a)$. For the scheme $(c)$,  $\tau/h$ can be even larger, up to $\sim 1$\,kHz, since it is driven by a single photon transition.

In the present model, and for subsequent results, we are not including the incoherent decay of level $\ket{1}$ which is generally present in real QCL structures, although it is especially connected to laser emission \cite{Faist:2013bo}. 
%
%
In the atomic simulator, such term can be made negligibly small, and will not be further considered. 


\section{Dynamics in simplified models} 

Focusing at the optimization of the laser emission rate, 
we consider the coherent coupling $\Omega$ and the decay rate $\gamma$ as the tunable parameters in the simulation, while the tunneling amplitude $\tau$ is taken as the overall energy and timescale reference. This last value for three-quantum-wells QCL active regions is typically 0.5~THz \cite{Faist:2013bo}, while $\Omega \simeq 200$~GHz and $\gamma \simeq 2.5$~THz \cite{Faist:1996} for a typical mid-infrared QCL. To maximize the laser emission rate, we aim to have at once a large population inversion and an efficient transport through the wells.

\subsection{Single-well toy model}
\label{4_levels} 

\begin{figure}[!b]
\centering
\includegraphics[width=0.6\columnwidth]{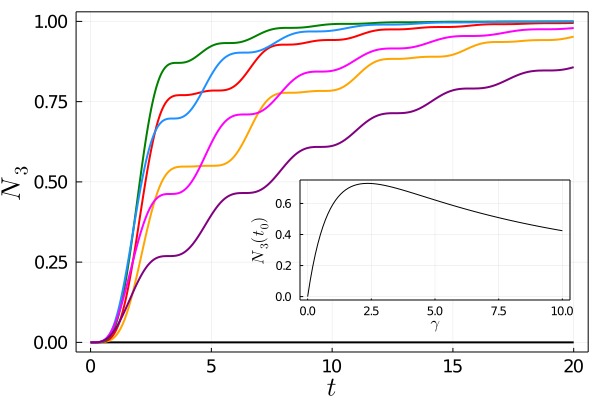}
\caption{Time evolution of $N_3$ for $\gamma=0$ (black),
$0.5$ (orange) $1$ (red), $2$ (green), $5$ (blue), $10$ (magenta) and $20$ (purple) in the 4-level model. 
The values of the parameters $\tau, \Omega$ are chosen to be $\tau=\Omega=1$; 
time is in units of $\hbar/\tau$. Inset: Values of $N_3$ at time $t_0=2.5$ as a function of $\gamma$ in the 4-level model.}
\label{fig1_and}
\end{figure}

Let our analysis start from the building block of the transport process, in which there are $3$ states in the generic well $j$. 
The more numerous are the particles in $\ket{3}$ levels, 
the more favoured is the transport. Particles 
are coming from the well $j-1$, and more precisely from the level $\ket{3}$ in the well $j-1$, 
and this is introduced in the model by having the population at the initial time $t=0$ entirely in the state $\ket{j-1,3}$ denoted, for the sake of brevity, 
by $\ket{0}$. 

Dropping the index $j$, we rewrite the Hamiltonian and the Lindbladian as: 
\begin{eqnarray}
\label{eq:Ham_toy}
H &=&\tau (c^\dag_{0} c_{1}+h.c.)+\Omega(c^\dag_{1} c_{2}+h.c.) \\
{\cal L}&=&\gamma L \rho L^\dag-\frac{\gamma}{2} \{ L^\dag L,\rho\}, 
\end{eqnarray}
with $L\equiv c^\dag_{3} c_{2}$. Furthermore, since $\tau$ sets the energy scale, we redefine $\Omega$ and $\gamma$ in units of $\tau$: in practice we set $\tau=1$.

Let us discuss now the competition taking place between the coherent coupling, $\propto \Omega$, 
and the decay term, $\propto \gamma$. 
Supposing for simplicity that $\Omega=1$, when $\gamma=0$, the population
moves from $\ket{0}$ to $\ket{1}$ and to $\ket{2}$, 
but no population ends in $\ket{3}$, thus no transport occurs. 
When $\gamma\neq0$ transport starts but  
if $\gamma$ becomes too large, the 
decay $\propto \gamma$ acts as an effective measurement and, due to the quantum Zeno effect \cite{Facchi:2008}, the population remains in the subspace spanned by $\ket{0}$, $\ket{1}$ and $\ket{2}$ 
for times increasing with $\gamma$. 
Therefore, 
at any given time $t_0$,
the number of particles in $\ket{3}$ increases 
with $\gamma$ for $\gamma$ smaller than an optimal value $\gamma_{opt}$, and then decreases upon further increasing $\gamma$ with $\gamma>\gamma_{opt}$. This behaviour is shown in 
Fig.~\ref{fig1_and}, where $N_3(t)=\langle c^\dag_{3} c_{3}\rangle$ denotes the number of particles 
in level $\ket{3}$ at time $t$. 
The initial state is chosen to be as $c_0^{\dag}\ket{vac}$, i.e. one particle 
in the level $\ket{0}$. 
The presence of an optimal value $\gamma_{opt}$ (which depends on $t_0$) is clearly seen in the inset. 
For $\Omega \neq 1$, the non-monotonic behaviour is still present, and if $\Omega>1$
(easier to implement with respect to the case $\Omega < 1$, see the discussion in the previous Section), $\gamma_{opt}$ increases with $\Omega$.

\subsection{Two wells with periodic boundary conditions}
\label{2W}
\begin{figure}[!htbp]
\centering
\includegraphics[width=0.3\textwidth]{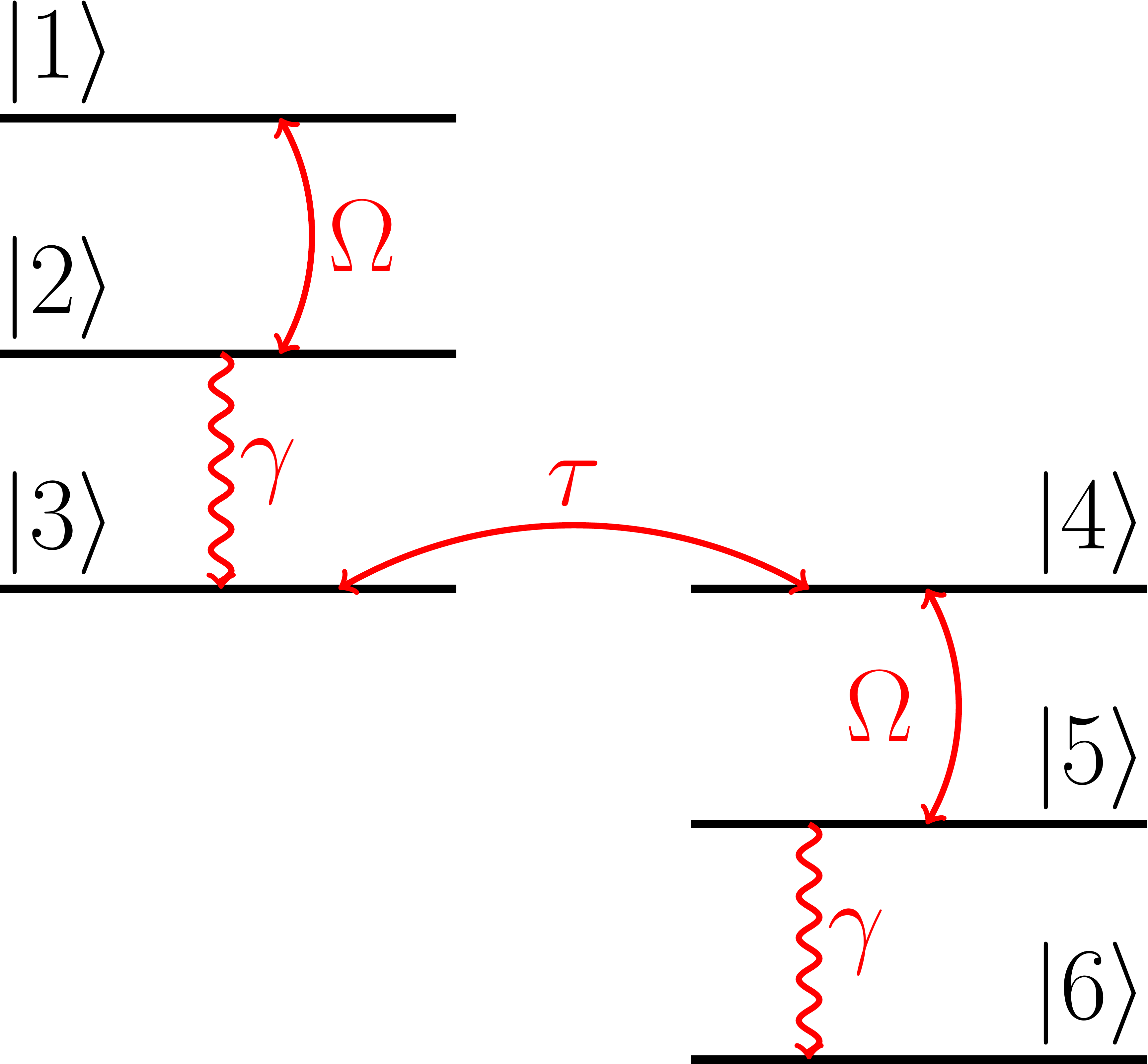}
\caption{Labelling of the levels for the two-wells model 
of Section \ref{2W} (notice that in literature the notation with the ket $\ket{1}$ in the bottom is often used):
$\tau$ is the tunneling between wells, $\Omega$ the coherent coupling on the laser transition, and
$\gamma$ the radiative decay of the lower laser level. Periodic boundary conditions imply that tunneling $\tau$ connects also $\ket{6}$ with $\ket{1}$ (not shown).} 
\label{fig0_and}
\end{figure}

The above discussion focused on two subsequent wells,
with the Lindbladian term acting to accumulate particles in the lowest level of the second well. Then, it is natural to ask what are the effects on the full cascade of wells, and how we can characterize the competition between the different terms producing a stationary state. Clearly, a complete answer to these issues may come from the proposed quantum simulation with cold atoms, however  here we provide a simple description and a first qualitative understanding from considering two subsequent wells with {\em periodic} boundary conditions \cite{Franckie:2018}. 
With the two wells labeled by $j$ and $j+1$, let denote now
the operators $c^{\dag}_{j-1,\alpha}$
by $c^{\dag}_\alpha$ with $\alpha=1,2,3$ and similarly   
$c^{\dag}_{j,\alpha}$ by $c^{\dag}_{\alpha+3}$, 
so that the operators $c_1,c_2,c_3$ refer to the fermionic operators destroying particles in the levels of the well $j-1$, and $c_4,c_5,c_6$ to the ones destroying particles in the levels of the well $j$. 
A schematic representation of this notation is in Fig. \ref{fig0_and}. 

We then obtain equation (\ref{eq:rho}) with 
\begin{equation}
H=\tau c^\dag_{3} c_{4} + \Omega(c^\dag_{1} c_{2} +
c^\dag_{4} c_{5})+ \tau  c^\dag_{6} c_{1} + h.c. 
\end{equation}
and ${\cal L}$ given by
\begin{equation}
      {\cal L}=\gamma \sum_{j=1,2} \Big(
      L_j \rho L_j^\dag-\frac{1}{2} \{ L_j^\dag L_j,\rho\} \Big),
\end{equation}
with $L_1=c^\dag_{3} c_{2}$ and $L_2=c^\dag_{6} c_{5}$. 

The main difference with the $4$-levels case analyzed in Section 
\ref{4_levels} is that a non-trivial stationary state is reached. With $N_i \equiv \langle c_i^\dag c_i \rangle$, the population of the lowest level in the second well, 
$N_6$, reaches an asymptotic value after a transient that is longer both for small and large values of $\gamma$. 

In the following considerations and figures we are considering as initial state 
$\rho(t=0)=\ket{1} \bra{1}$, i.e., all the population in the level $\ket{1}$. Notice that, our periodic boundary conditions imply that
the asymptotic values of quantities will be the same in the two wells, e.g. $N_1^{asympt}=N_4^{asympt}$, $N_2^{asympt}=N_5^{asympt}$ and so on (labelling with $^{asympt}$ the quantities for $t \to \infty$). 

Actually, we use 6 levels  to clearly illustrate 
the motion of particles between the wells. A minimal model could restrict to a single well, i.e. levels $\ket{1}, \ket{2}, \ket{3}$, and implement the periodic boundary conditions by having tunneling between $\ket{1}$ and $\ket{3}$. For the cases discussed below, we verified that the two models yield identical results for the asymptotic values.

To assess the performance of the modeled QCL, our main interest is upon the following quantities:
\begin{itemize}

\item the expectation value $I$ of the current operator
$\hat{I}=- i \,
(c^\dag_3 c_4 - c^\dag_4 c_3 )$, given by $I=- 2 \, 
\mathrm{Im} \left[\rho_{43}\right]$ (with this choice, for the considered initial condition, $I>0$);

\item the population inversion (e.g., in the second well):
\begin{equation}
\Delta N_{inv}(t)=N_4(t)-N_5(t), 
\end{equation}
so that the population is truly inverted on the laser transition for $\Delta N_{inv}>0$. As a good proxy of the laser emission rate, we will take $\Delta N_{inv} \Omega^2$ \cite{Demtroder2014}. 
\end{itemize}

The goal of the following discussion is to summarize the dependence of the asymptotic values of the above quantities upon $\Omega$ and $\gamma$, and to point out the occurrence of optimal values.

As in previous Section, we start with the case $\Omega=\tau$ and then generalize to $\Omega \neq \tau$. However, it is interesting to observe that certain relations for asymptotic values hold in both cases. First we get 
\begin{equation}
|\rho_{43}|^{asympt}= \frac{I^{asympt}}{2}    
\label{gen1}    
\end{equation}
(actually valid at {\it any} time). Moreover we obtain 
\begin{eqnarray}
\gamma \, N_5^{asympt} &=& \tau I^{asympt}   
\label{gen2}    \\
\tau \, (\mathrm{Im}[\rho_{43}])^{asympt} &=& \Omega \, (\mathrm{Im}[\rho_{54}])^{asympt}.   
\label{gen3}    
\end{eqnarray}
Eqs. (\ref{gen1}) and (\ref{gen2}) are particularly useful since they relate the current to the coherence, and the decay rate
to the current via the occupation of the intermediate level, respectively.

\subsubsection{\underline{$\Omega=\tau$}}

Let our analysis start from the case $\Omega=\tau$. Solving the Lindblad equation, it clearly appears that the (desired) inversion takes place. Typical plots of the time dependence of the different quantities are reported in Appendix, to which we defer for further details. First, we observe that, due to symmetry 
$\tau \leftrightarrow \Omega$, we obtain  
\begin{equation}
    N_1^{asympt}=N_3^{asympt}; \,\,\, \,\,\, \,\,\,
\,\,\, \,\,\, \,\,\,
N_4^{asympt}=N_6^{asympt}.
\label{symm}
\end{equation}
Since for the symmetry induced by the periodic boundary conditions it is $N_1^{asympt}=N_4^{asympt}$, $N_2^{asympt}=N_5^{asympt}$, then it follows that 
$N_4-N_3 \to 0$ for large time, as it can be verified. 
One could add a Lindbladian term of the form  $\tilde{{\cal L}}=
\tilde{\gamma} \sum_{j=1,2} \Big(
      \tilde{L}_j \rho \tilde{L}_j^\dag-\frac{1}{2} \{ \tilde{L}_j^\dag \tilde{L}_j,\rho\} \Big)$ 
with $\tilde{L}_1=c^\dag_{3} c_{1}$ and $\tilde{L}_2=c^\dag_{6} c_{4}$, i.e., adding a dissipative term incoherently 
transferring atoms from the highest to the lowest
level of each well, that, in the atomic simulators described in the previous Section, is typically larger than the incoherent decay term from $\ket{1}$ to $\ket{2}$). In that case symmetry given by Eqs. (\ref{symm}) no longer holds.  

\begin{figure}[!htbp]
\centering
\includegraphics[width=0.6\columnwidth]{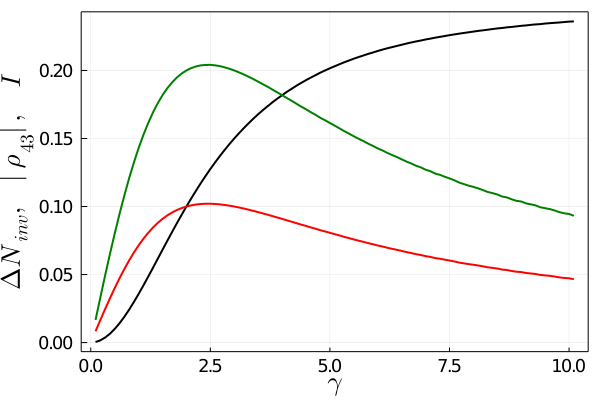}
\caption{Long-time values of $\Delta N_{inv}$ (black), 
$|\rho_{43}|$ (red) and $I$ (green) as a function of the decay parameter $\gamma$ (with $\tau=\Omega=1$).}
\label{fig2_and}
\end{figure}

Our results are shown in Fig.~\ref{fig2_and}, where we plot $\Delta N_{inv}^{asympt}$, 
$|\rho_{43}|^{asympt}$ and $I^{asympt}$ as a function 
of $\gamma$. 

In Fig. \ref{fig2_and} 
we find a non-monotonic behaviour for the transport quantities, i.e. the current $I$ and of the coherence $|\rho_{43}|$, for which the optimal values of $\gamma$ are finite. This results from the combined presence of the tunneling and the dissipative terms: if $\gamma$ is vanishingly small the population does not reach the lowest level, while for large $\gamma$ the coherence and the tunneling between wells are suppressed.

At variance, the population inversion $\Delta N_{inv}$ and the stimulated emission rate grow with $\gamma$, the former reaching the value $0.25$ in the large $\gamma$ limit. The reason is that for large values of $\gamma$ the occupation of the intermediate levels 
$\ket{2}$ and $\ket{5}$ is increasingly suppressed, so that $N_2^{asympt}=N_5^{asympt} \to 0$ for $\gamma
 \to \infty$. Then, due to eq. (\ref{symm}) all the other four values go the same value, which for the normalization has to be 
 $N_1^{asympt}=N_3^{asympt}=N_4^{asympt}=N_6^{asympt}=
 1/4$.

\begin{figure}[!htbp]
\centering
\includegraphics[width=0.45\columnwidth]{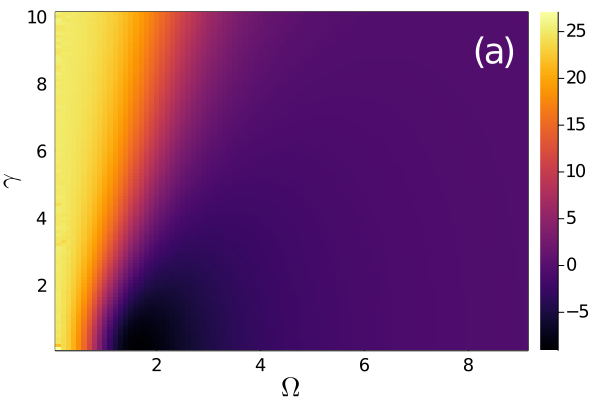} 
\includegraphics[width=0.45\columnwidth]{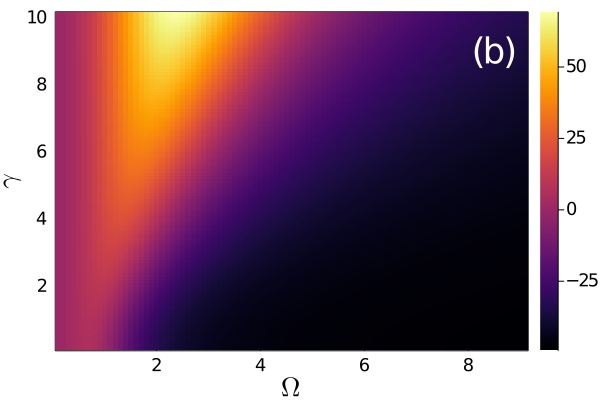} \\
\includegraphics[width=0.45\columnwidth]{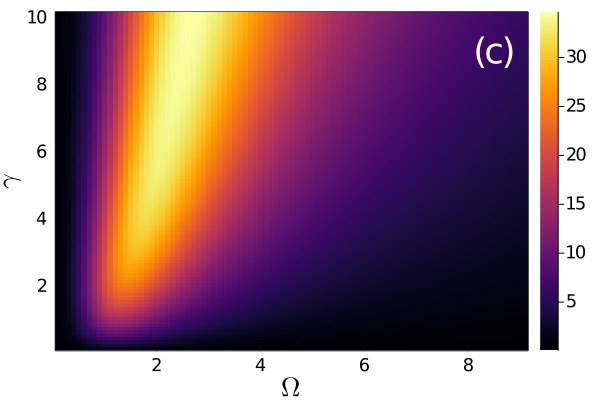} 
\includegraphics[width=0.45\columnwidth]{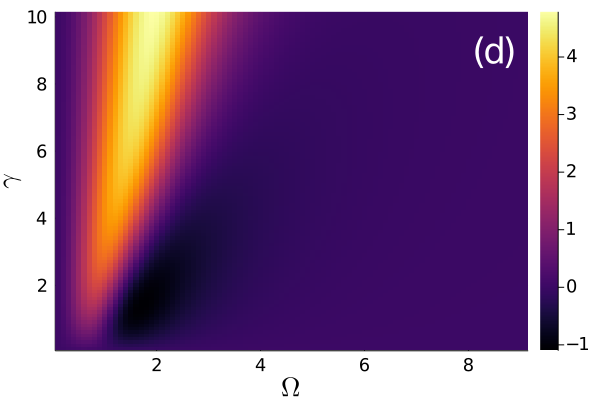} 
\caption{Asymptotic values of: population inversion $\Delta N_{inv}$ (a), laser emission rate $\Omega^2\Delta N_{inv}$ (b), current $I$ (c), and $I \Delta N_{inv}$ (d). For clarity, we show these quantities multiplied by 100 
(moreover, $\tau=1$).}
\label{fig:contours}
\end{figure}

\subsubsection{\underline{$\Omega \neq \tau$}}

The first observation is that the population inversion increases with lowering $\Omega$, so that its maximum value occurs for $\Omega =0$, as shown in Fig. \ref{fig:contours} (a).
This is quite reasonable as for low values of the coherent coupling $\Omega$ the population of the intermediate states $\ket{2}$ and $\ket{5}$ vanishes. Actually, as $\Omega$ increases the inversion itself disappears,
i.e., $\Delta N_{inv}$ turns negative at a specific value $\Omega_{0}$. For $\Omega > \Omega_0$, population inversion does not take place, it reaches a negative minimum and then grows again, remaining negative and vanishing for large $\Omega$.

\begin{figure}[!htbp]
\centering
\includegraphics[width=0.55\columnwidth]{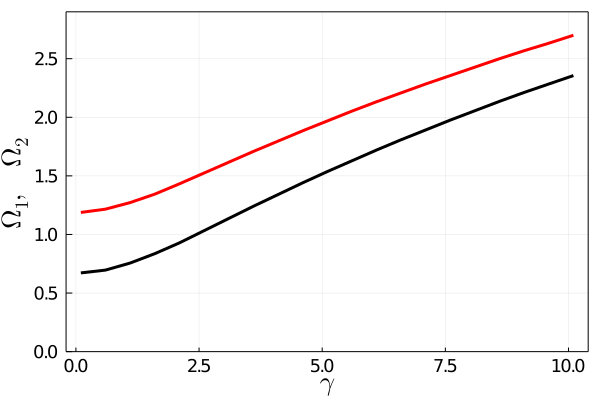}
\caption{Value of the coherent coupling $\Omega_1$ ($\Omega_2)$, black (red) line, that maximizes the stimulated emission rate (the current), as a function of $\gamma$,
for $\tau=1$.}
\label{fig:Omega12}
\end{figure}

Related to the population inversion, the stimulated emission rate $\Delta N_{inv} \Omega^2$ 
displays a non-monotonic behaviour, with a maximum for $\Omega = \Omega_1$ for any given value of $\gamma$, see Fig. \ref{fig:contours} (b). To maximize the laser emission, the coherent coupling $\Omega$ must be chosen equal to $\Omega_1$: for small $\Omega$, the probability amplitude is suppressed, while for large $\Omega$ the population is not inverted and photon gain yields to photon absorption. The optimum value $\Omega_1$ is approximately linear with $\gamma$ as shown in 
Fig. \ref{fig:Omega12}.

As for the transport properties, the current $I$ also displays a non-monotonic behaviour, with a maximum for $\Omega = \Omega_2$, approximately linearly with $\gamma$ and also shown in Fig.~\ref{fig:Omega12}. 

We conclude by noticing that there is not a single figure of merit for the operation of a QCL. 
Optimizing the current, hence the transport of electrons across the QCL heterostructure, is certainly beneficial to the laser efficiency, but it is not the most appropriate figure of merit if we aim to the largest laser power: indeed, current flows also in a configuration of non-inverted population ($\Delta N_{inv}<0$) where the laser emission is absent. 
Reasonably, as far as the laser power is concerned, the most sensible figure of merit is the rate of stimulated emission. On the other hand, it is known that QCL linewidth depend upon the ratio of radiative to non-radiative decay of the upper laser level \cite{Faist:2013bo}, i.e. $\ket{1}$ and $\ket{4}$. Due to the absence of non-radiative decay, our model is unfit to optimize e.g. the spectral performance.
An additional goal in QCL design, not addressed within our model, is to achieve the highest possible operating temperature \cite{Franckie:2018}. 

With the aim of comparing the simulation results with real QCLs operation, we remark that with respect to the typical QCL parameters reported above, i.e. $\tau \simeq 0.5$ THz, $\Omega \simeq 0.4\tau$ and $\gamma \simeq 5\tau$, we find that both the laser emission rate and the current would improve with larger values of $\Omega$, approximately equal to $1.5 \tau$ and $2 \tau$, respectively. This shows a simple, first example of the potential of this simplified transport model to guide QCL design parameters, suggesting that a full quantum simulation with cold atoms, including explicitly interaction effects, will provide optimized parameters for future improved lasers. 

\section{Summary} 

We have proposed a cold-atom platform to simulate key features of electron transport in QCLs, that are deemed to be essential to understand and improve the performance of current devices. Specifically, we have described a QCL simulator to be realized with an 
ultracold Fermi gas 
trapped in a one-dimensional lattice in presence of a tunable linear magnetic field gradient, detailing the simulation engineering of the individual processes that electrons undergo in their transport along the QCL heterostructure. We argued that the atom-based simulator can be key to investigate aspects that are hardly dealt with present-day computational tools, such as electron-electron interactions and disorder, by exploiting Feshbach resonances and by appropriately shaping the trapping laser configuration, respectively.

In dealing with the simulation of electron transport in QCL heterostructures, the relevant parameters are the hopping coefficients, the inter-well tunneling $\tau$ and the intra-well coherent transition strength $\Omega$. They compete with the parameter $\gamma$, i.e. the strength of the incoherent decay present in the Lindbladian term ${\cal L}$. The parameters $\tau$, $\Omega$ and $\gamma$ can be varied within different schemes, which we discussed providing realistic estimates. 

Two simplified single-particle models have been numerically studied to guide the setup of the cold-atom experiment. The first model has a single well fed by particles coming from the adjacent well, while in the second model two wells are present with periodic boundary conditions. In the latter, asymptotic values are reached after a transient. In both cases, competition between the hopping terms $\tau, \Omega$ and the decay term $\gamma$ is evidenced, together with the presence of optimal values of $\gamma$ for the emission rate, the transport, and the coherence. In the former, single-well model, the ratio $\Omega/\tau$ does not play a qualitative role, and it merely shifts the optimal value of $\gamma$. However, this is not the case in the more realistic two-wells model, since the presence of population inversion depends on the ratio $\Omega/\tau$. 

Despite their toy-model nature, the two proposed models already display significant features, such as the competition between the hopping and the incoherent decay, the presence of non-monotonic behaviours for different relevant quantities (including the current) and the dependence of the population inversion on the model parameters. 

Important effects that are not included in the discussed models, and cannot easily be estimated by numerical simulations, can be added in a tunable way in the proposed quantum simulator, related to controllable disorder and interactions, emulating screened Coulomb repulsion among electrons in the structure. Interactions can provide an additional source of decoherence, but also a mechanism for thermalization of the longitudinal photon-assisted motion into transverse degrees of freedom. The present proposal can also be extended to include interactions with additional transverse degrees of freedom, which would be of importance in view of quantum simulations of realistic QCL structures. 

We consider this proposal the first step towards a quantum-assisted optimization of complex heterostructured devices via analog atomic simulations, providing a clear-cut example of how quantum simulators may impact on real-life technology. 

\section*{Acknowledgements} 

We thank F.~Benatti, J.~Faist, and M.~Francki\'e for useful discussions. 

The authors acknowledge financial support by the European Union’s Horizon 2020 Research and Innovation Programme with the Qombs Project [FET Flagship on Quantum Technologies grant n. 820419] ``Quantum simulation and entanglement engineering in quantum cascade laser frequency combs''.

\section*{Appendix: Detailed time dynamics 
of the two-well model}
\label{appendix}
In this Appendix we provide further details on the dynamics of the model discussed in Section \ref{2W}. 

We consider the case in which $\tau=1$ and $\Omega$ is fixed while $\gamma$ varies. 
Regarding the populations $N_i$, interestingly, 
for $\Omega \neq 1$ a non-monotonic behaviour of $N_6^{asympt}$ versus $\gamma$ arises. More in general, the coherences also display a non-monotonic behaviour with $\gamma$, and of course vanish for large $\gamma$. Due to Eq. (\ref{gen1}), the same behaviour is exhibited by $I$. At variance, $N_4$ has a monotonic behavior, unlike the population inversion $\Delta N_{inv}$ which becomes negative for large enough $\gamma$, then reduces back towards zero, 
then becoming negative and, again, assumes a non-monotonic behaviour vanishing for large $\gamma$.

\begin{figure}[!htbp]
\centering
\includegraphics[width=0.6\columnwidth]{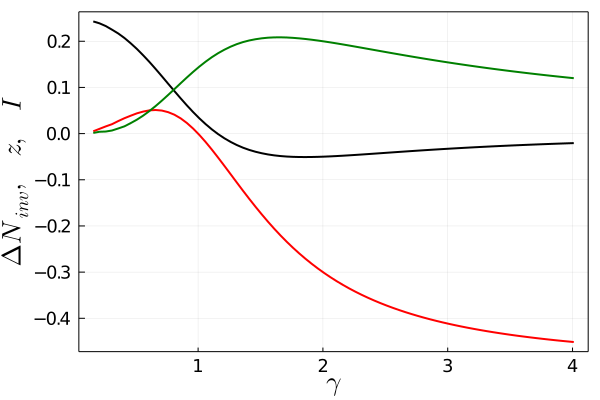}
\caption{Long-time values of $\Delta N_{inv}$ (black), 
$z$ (red) and $I$ (green) as a function of $\gamma$ 
for $\tau=1$ and $\Omega=\gamma$. 
}
\label{fig4_and}
\end{figure}

\begin{figure}[!htbp]
\centering
\includegraphics[width=0.6\columnwidth]{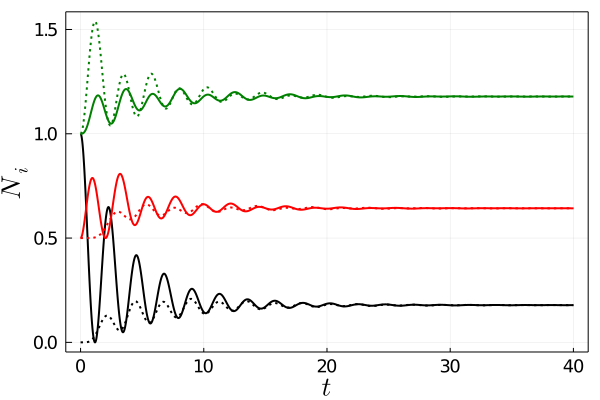}
\caption{Time evolution of the populations $N_i$ for $\tau=\Omega=\gamma=1$. $N_1$ ($N_4$): solid (dotted) black line; $N_2$ ($N_5$): solid (dotted) red line; $N_3$ ($N_6$): solid (dotted) green line. For clarity, $N_2$, $N_5$ are shifted up by 0.5, and $N_3$, $N_6$ by 1.}
\label{fig:populations_gamma1}
\end{figure}

\begin{figure}[!htbp]
\centering
\includegraphics[width=0.6\columnwidth]{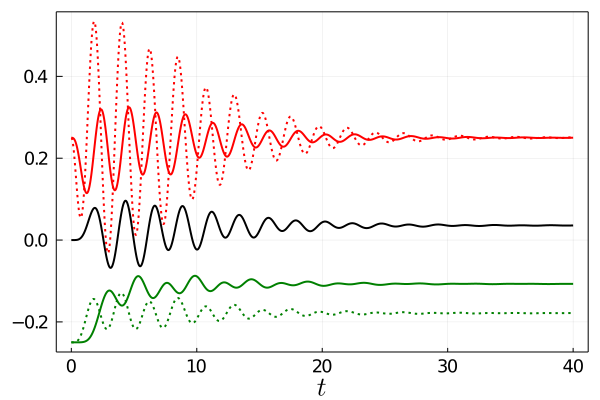}
\caption{
Time evolution of $\Delta N_{inv}=N_4-N_5$ 
(solid black line), $z$ and  $\dot z$ (solid and dashed red lines), $I$ and $|\rho_{43}|$ (solid and dashed green lines), for the same parameters of Fig. 
\ref{fig4_and}, i.e. $\tau=\Omega=\gamma=1$. For clarity, red (green) lines are shifted up (down) by 0.25.
}
\label{fig:DN_z_I_gamma1}
\end{figure}

\begin{figure}[!htbp]
\centering
\includegraphics[width=0.6\columnwidth]{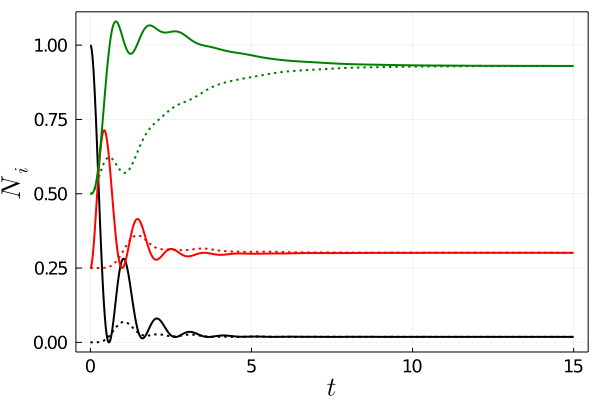}
\caption{
Time evolution of the populations $N_i$ for $\tau=1$ and $\Omega=\gamma=3$. $N_1$ ($N_4$): solid (dotted) black line; $N_2$ ($N_5$): solid (dotted) red line; $N_3$ ($N_6$): solid (dotted) green line. For clarity, red (green) lines are shifted up by 0.25 (0.5).}
\label{fig:populations_omega3gamma3}
\end{figure}

In Fig. \ref{fig4_and} we plot the asymptotic values 
of $\Delta N_{inv}$, $z$ and $I$ as a function of $\gamma$ 
for $\tau$ fixed and $\Omega=\gamma$.
In Fig. \ref{fig:populations_gamma1} we plot the time dependence 
of the populations $N_i$ ($i=1,\cdots,6$) for $\tau=\Omega$ and a particular value of $\gamma$. It holds that $N_1^{asympt}=N_3^{asympt}$ and $N_4^{asympt}=N_6^{asympt}$. We verified that if a incoherent Lindbladian decay is added between levels 
$\ket{1}-\ket{3}$ and $\ket{4}-\ket{6}$, then it is not longer true that $N_1^{asympt}=N_3^{asympt}$ and $N_4^{asympt}=N_6^{asympt}$, but the relations  $N_1^{asympt}=N_4^{asympt}$ and $N_3^{asympt}=N_6^{asympt}$ continue to hold. 
In Fig. \ref{fig:DN_z_I_gamma1} we plot the time evolution of $\Delta N_{inv}=N_4-N_5$, $z=N_4-N_3$, $\dot z$, $I$ and $|\rho_{43}|$ for $\tau=\Omega=\gamma=1$. In Fig. \ref{fig:populations_omega3gamma3} we plot the time dependence 
of the different quantities of interest for the same value 
of $\gamma$ used in Fig. \ref{fig:DN_z_I_gamma1}: 
$\Delta N_{inv}$, $z$, $\dot{z}$, $I$, $|\rho_{43}|$, where $z(t) = N_4(t)-N_3(t)$ is the imbalance between the highest level of the second well and the lowest of the first, and $\dot z$ is
the rate of population transfer between the two wells. 
For this particular value of $\gamma$, asymptotically, a population inversion condition is reached. 
Due to having reached the asymptotic stare, it is $\dot{z} \to 0$ for $t\to \infty$. In agreement with the fact that $N_1^{asympt}=N_3^{asympt}$ and $N_4^{asympt}=N_6^{asympt}$ we find that $z \to 0$ for large times. 

Fig. \ref{fig:populations_omega3gamma3} gives the time dynamics of the populations $N_i$ for a case with $\tau < \Omega$ 
($\tau=1$, $\Omega=3$, $\gamma=3$). We obtain that $N_1^{asympt} \neq N_3^{asympt}$ and that the inversion 
does not take place, as depicted in Fig. \ref{fig:populations_omega3gamma3} where also $z$, $\dot{z}$, 
$I$ and $|\rho_{43}|$ are plotted. 
Figs. \ref{fig:populations_omega05} and \ref{fig:DN_z_I_omega05} refer to a case featuring $\tau > \Omega$: there, we see inversion. 

\begin{figure}[!htbp]
\centering
\includegraphics[width=0.6\columnwidth]{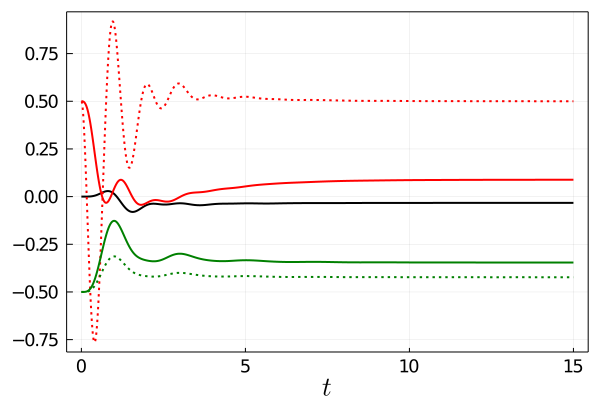}
\caption{
Time evolution of $\Delta N_{inv}=N_4-N_5$ 
(solid black line), $z$ and  $\dot z$ (solid and dashed red lines), $I$ and $|\rho_{43}|$ (solid and dashed green lines), for the same parameters of Fig. \ref{fig:populations_omega3gamma3}, i.e. $\tau=1$ and 
$\Omega=\gamma=3$. For clarity, red (green) lines are shifted up (down) by 0.5.}
\label{fig:DN_z_I_omega3gamma3}
\end{figure}

\begin{figure}[!htbp]
\centering
\includegraphics[width=0.6\columnwidth]{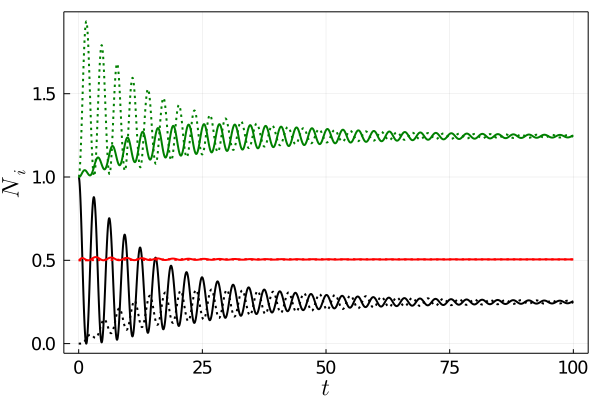}
\caption{Time evolution of the populations $N_i$ for $\tau=1$ and $\Omega=0.4$, with  
$\gamma=5$. $N_1$ ($N_4$): solid (dotted) black line; $N_2$ ($N_5$): solid (dotted) red line; $N_3 (N_6)$: solid (dotted) green line. For clarity, red (green) lines are shifted up by 0.5 (1.0).}
\label{fig:populations_omega05}
\end{figure}

\begin{figure}[!htbp]
\centering
\includegraphics[width=0.6\columnwidth]{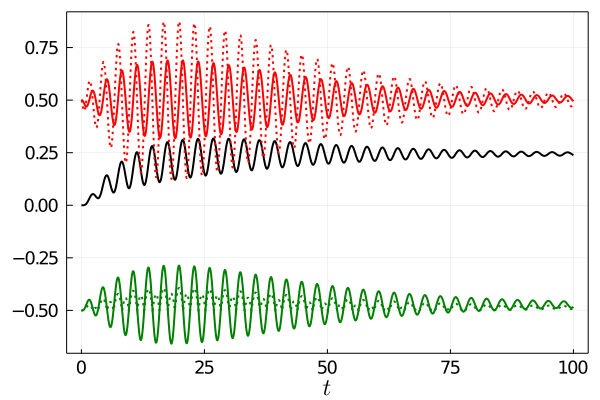}
\caption{
Time evolution of $\Delta N_{inv}=N_4-N_5$ 
(solid black line), $z$ and  $\dot z$ (solid and dashed red lines), $I$ and $|\rho_{43}|$ (solid and dashed green lines), for the same parameters of Fig. \ref{fig:populations_omega05}, i.e. $\tau=1$, 
$\Omega=0.4$ and $\gamma=5$. For clarity, red (green) lines are shifted up (down) by 0.5.}
\label{fig:DN_z_I_omega05}
\end{figure}

\clearpage

\bibliographystyle{ieeetr_CA}
\bibliography{references_r1}




\end{document}